\documentclass[review]{elsarticle}

\usepackage{amsmath}
\usepackage{multirow}

\usepackage{lineno,hyperref}
\modulolinenumbers[5]

\usepackage{cleveref}

\journal{arXiv.org}

\biboptions{authoryear}

\def \RSA		{\textit{Routing and Spectrum Allocation}}
\def \cplex     {\textsc{Cplex}}
\def \CG		{\textit{Column Generation}}
\def \NCG		{\textit{Nested Column Generation}}

\def \BnB       {\textit{Branch-and-Bound}}
\def \BnP		{\textit{Branch-and-Price}}
\def \MP		{\textit{Master Problem}}
\def \RMP		{\textit{Restricted} \MP{}}
\def \PP		{\textit{Pricing Problem}}
\def \PPs		{\textit{Pricing Problems}}

\def \EON		{\textit{Elastic Optical Network}}

\def \ILP		{\textit{Integer Linear Program (ILP)}}
\def \ILPing	{\textit{Integer Linear Programming}}

\def \lp		{\textit{Lightpath}}

\def \conf		{\textit{Configuration}}
\def \confs		{\textit{Configurations}}

\def \LCGS      {\textsc{L\_CG}}

\def \CCG       {\textsc{C\_CG}}

\def \CNCG      {\textsc{C\_NCG}}

\def \zILPstar			{z^{\star}_{ILP}}
\def \zILPtilda			{{\tilde z}_{ILP}}
\def \zLPstar			{z_{LP}^{\star}}

\def \rcILPstar			{rc^{\star}_{ILP}}
\def \rcILPtilda		{{\tilde {rc}}_{ILP}}
\def \rcLPstar			{rc_{LP}^{\star}}

\def \betap	       	{\beta_{\pi}}

\def \akc 			{a_k^c}
\def \blsprime 		{b_\ell^{s'}}
\def \blsc 			{b_\ell^{sc}}
\def \muk 			{\mu_k}
\def \mulsprime 	{\mu_{\ell s'}}

\def \ak 			{a_k}

\def \nuk 			{\nu_k}
\def \nul 			{\nu_\ell}

\def \deltapl       {\delta_{\pi}^{\ell}}

\def \ysd			{y_{sd}}
\def \asdc			{a_{sd}^c}

\begin{document}

\begin{frontmatter}

\title{Nested Column Generation decomposition for solving \\ the Routing and Spectrum Allocation problem \\ in Elastic Optical Networks}

\author{Julian Enoch\corref{mycorrespondingauthor}}
\cortext[mycorrespondingauthor]{Corresponding author}
\ead{julian.enoch@utdallas.edu}

\address{Computer Science and Software Engineering \\ Concordia University}


\begin{abstract}
With the continued growth of Internet traffic, and the scarcity of the optical spectrum, there is a continuous need to optimize the usage of this resource. In the process of provisioning elastic optical networks using the flexible frequency grid paradigm, telecommunication operators must deal with a combinatorial optimization problem that is NP-complete namely the \RSA{} (RSA) problem.

Following on our previous study, where we used \ILPing{}, and proposed a \CG{} algorithm based on a \lp{} decomposition, which proved to be the most efficient so far, we now consider the traditional \conf{} decomposition that has been studied in other works in the past. In the process, we created an new mathematical model using two variable sets instead of a single variable set. Equally important, we independently rediscovered the \NCG{} technique, and we used it to propose an algorithm that led to a considerable improvement on the previous algorithms that use the same \conf{} decomposition. When compared to the latest such existing study, our algorithm achieved an accuracy gap of 1\% as opposed to 14.3\% for the previous study, and a running time two orders of magnitude faster on average.

\end{abstract}

\begin{keyword}
Nested Column Generation (NCG),
Integer Linear Programming (ILP),
Large Scale Optimization,
Combinatorics,
Elastic Optical Networks (EON),
Routing and Spectrum Allocation (RSA)
\end{keyword}

\end{frontmatter}


\section{Introduction} \label{chap:introdcution}

The core networks of Internet (called also long-haul networks) are based on optical technology and their links are made of optical fibers. As a way to increase the capacity, these networks implement Wavelength Division Multiplexing (WDM), where the idea is to transmit data simultaneously at multiple carrier optical frequencies. Traditionally, the multiplexing was done according to a Coarse WDM frequency grid (called fixed grid) where the carrier wavelength granularity is 50 GHz. As of 2012, the International Telecommunication Union issued a new recommendation that defines a flexible Dense WDM frequency grid with a finer frequency granularity of 12.5 GHz called a frequency slot \citep{iut12}. This evolution has allowed the emergence of Elastic Optical Networks.

The main phase in the process of designing an Elastic Optical Network is concerned with traffic provisioning, which consists of assigning a routing path, and allocating an optical spectrum range to each traffic demand. The spectrum range corresponds to a number of contiguous frequency slots. This provisioning problem is called the \RSA{} (RSA) problem.

The RSA optimization problem has been shown to be NP-hard by \cite{shir13}. 
Among the existing algorithms, heuristics are in general fast but they do not provide any guarantee on the quality of the solutions. Therefore, in order to solve it \textit{exactly} (i.e., with a calculated accuracy), we need to employ \ILPing{} (ILP). While ILP compact formulations (e.g., link-based) have a polynomial number of variables, they do not scale well when the problem instances reach a real-life size. Therefore, we consider decomposition formulations.

\subsection{Existing studies}

Various mathematical models have already been proposed for solving the RSA problem, including several column generation models, however with different decomposition schemes, and different solution processes. All classical ILP models are not scalable and can only solve very small data instances, i.e., with few nodes and with a spectrum made of a very limited number of slices.

\cite{ruiICTON13} proposed a first column generation model, with the minimization of the number of denied demands and the amount of non-served bit-rate. They are able to solve data instances with up to 96 slots, and an overall demand distributed over a set of 180 node pairs in the Spain network (21 nodes).

\cite{kli14_ICTON} improved the previous formulation with the use of valid inequalities (cuts), but did not go significantly farther. 

\cite{kli15_IDEAL} reformulated the RSA problem as a mixed-integer program and solved it using a branch-and-price algorithm. In order to enhance the performance of their algorithm, a simulated annealing-based heuristic was added. 

In a recent attempt to exactly solve large-size instances of RSA problem, \cite{kli16} proposed a branch-and-price algorithm. However, the resulting algorithm is not an exact algorithm and the Linear Program (LP) value is not a valid bound to assess the quality of the ILP solutions, as the authors use pre-computed lightpaths, and did not consider all possible lightpaths.

In our previous study \citep{enoch18_inoc}, we proposed an ILP formulation based on a \lp{} decomposition, and we solved it using \CG{} technique. This \lp{} decomposition proved to be very efficient compared to previous algorithms. But, as we scaled up our problem instances, we sensed the limits of its efficiency. In fact, the fine-granularity of the \lp{} decomposition makes it that the final ILP in the \CG{} process contains a very high number of columns, and can take very long to be solved. Consequently, we decided to explore a different decomposition based on a column definition as a set of lightpaths referred to as a \conf{} of lightpaths.

The \conf{} decomposition has been previously explored in \citep{jau16_ICTON} using \CG{} and combining two algorithms to solve the \PP{}, one of which is a \textit{Path-based} ILP that is restricted to a pre-computed subset of paths and generates feasible \textit{Configurations} quickly, and the other one is an optimal \textit{Link-based} ILP used at the end of \CG{} to guarantee the optimality of the \PP{}. However, this solution suffers from two issues. The pre-computation for the \textit{Path-based} \PP{} is complex, and the \textit{Link-based} \PP{} is inefficient.


In the current study, we introduced a new formulation for the \conf{} decomposition, and more importantly, we independently rediscovered the \NCG{} technique that expands on the traditional \CG{} approach. The inspiration behind the \NCG{} technique was to have only one \textit{Path-based} ILP for the \PP{}, and solve it optimally and efficiently, which leads to an overall efficient algorithm.


Throughout this document, we use the following acronyms to refer to different algorithms:

\begin{itemize}
	\item \LCGS{} refers to the \CG{} algorithm using a \lp{} decomposition introduced by the same author of this work in \citep{enoch18_inoc};
	\item \CCG{} refers to a \CG{} algorithm using a \conf{} decomposition introduced by \cite{jau16_ICTON};
	\item \CNCG{} refers to the new \NCG{} algorithm using a \conf{} decomposition introduced in this work;
\end{itemize}

\section{RSA Decomposition Model}
\label{sec:model_ejor}


\subsection{Statement of the RSA Problem}

We consider an \EON{} and we represent its topology by an undirected graph $G=(V,L)$ with node set $V$ and link set $L$.
The bandwidth spectrum of the optical network is sliced into a set $S$ of frequency slots of width 12.5GHz.
The traffic is defined by a set $K$ of optical connections where each connection $k \in K$ is defined by a source $s_k$, a destination $d_k$ and a bandwidth demand $D_k$, expressed in terms of a number of slots.

Given an \EON{}, and a traffic demand matrix between its nodes, the RSA problem consists of finding optical channels for the traffic requests, as to maximize the overall throughput. For a given traffic request $k \in K$ from source node $s_k$ to destination node $d_k$, with spectrum demand $D_k$, an optical channel (or lightpath) is a combination of a routing path $\Pi_k$ from node $s_k$ to node $d_k$, and an optical slice that is composed of a contiguous set of spectrum slots and that is continuous throughout the routing path $\Pi_k$. An optical slice is - by definition - a contiguous set of frequency slots, thus it can be characterized (in addition to its width $D_k$) by its low-end spectrum slot which we refer to as a \textit{starting slot}.


\subsection{Decomposition Model}

\begin{figure}[ht]
	\centering
	\includegraphics[width=1.\linewidth]{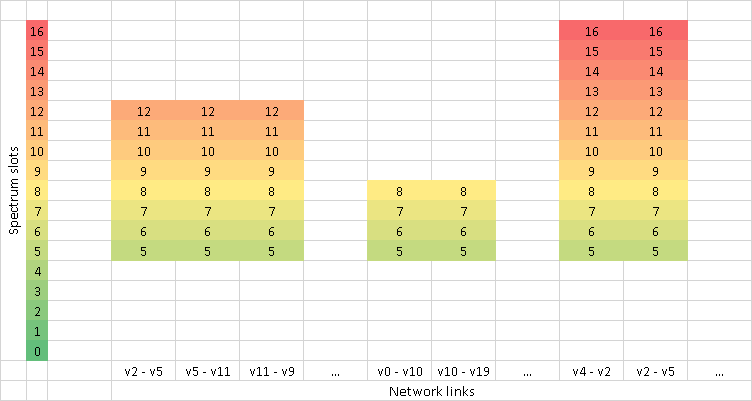}
	\caption{\conf{} illustration - The first request connects node-pair $\{v_2, v_9\}$. It is routed using three links. Its demand is 8 slots. This configuration contains three lightpaths all of which start at slot 5.}
	\label{fig:config}
\end{figure}

The ILP model is based on a \conf{} decomposition. We define a \conf{} as a set of link-disjoint lightpaths provisioning distinct requests and having the same starting slot. 
\Cref{fig:config} shows an example of a \conf{} of three lightpaths.

We have experimented with the mathematical model proposed in \citep{jau16_ICTON}, but we observed that it performed poorly both in terms of the quality of the solution, and in terms of the computation time. The intuition behind this is that the \conf{} columns are relatively large which makes LP relaxation take longer to converge to its optimal value ($\zLPstar{}$). This also leads to a high contention among the \confs{} during the ILP \BnB{} phase, which produces an integral solution ($\zILPtilda{}$) with a relatively large accuracy gap.

Therefore, we created a new mathematical model that employs two sets of variables instead of only one, and that proved to be easier to solve.

Let $C$ be the set of all possible \textit{Configurations}. Each \conf{} $c \in C$ is defined with the help of the following parameters:
\begin{itemize}
	\item $\akc$: indicates if a request $k$ is provisioned by \conf{} $c$.
	\item $\blsc$: indicates if one of the requests provisioned in \conf{} $c$ is routed on link $\ell$ while assigned slot $s$ as part of its spectrum slice.
\end{itemize}

The two sets of decision variables are as follows:

\begin{itemize}
	\item $z_c$ is a binary variable that indicates if \conf{} $c$ is selected in the solution.
	\item $y_k$ is an intermediate binary variable that indicates if a request $k$ is granted in the solution, and can be expressed in terms of $z_c$ as follows:
\begin{equation}
            y_k = \sum_{c \in C} \akc z_c \qquad k \in K. \label{eq:yk_definition} 
\end{equation}
\end{itemize}

The objective function consists of maximizing the throughput and is written in terms of $y_k$ as follows:

\begin{equation}
z \quad = \quad \max \quad  \sum_{k \in K} D_k y_k  \label{eq:objective_ejor} 
\end{equation}

subject to:

\begin{alignat}{2}
  &  y_k \leq \sum_{c \in C} \akc z_c \qquad    && k \in K, \label{eq:one_config_per_k_ejor}
\\&  \sum_{c \in C} \blsc z_c \leq 1 \qquad     && \ell \in L, s \in S, \label{eq:G_config_disjointness_ejor}
\\&  z_c \in \{0, 1\} \qquad                    && c \in C, \label{eq:domain_zc_ejor} 
\\&  0 \leq y_k \leq 1 \qquad                   && k \in K. \label{eq:domain_yk} 
\end{alignat}

Constraints \eqref{eq:one_config_per_k_ejor} means that a request $k$ is granted only if it is provisioned in at least one of the \confs{} $z_c$.
Constraints \eqref{eq:G_config_disjointness_ejor} make sure that each slot on each fiber link is not used by more than one \conf{} $z_c$.
The integrality of variables $y_k$ is guaranteed by the combination of \eqref{eq:objective_ejor} and \eqref{eq:one_config_per_k_ejor}, and needs not to be explicitly enforced in \eqref{eq:domain_yk}.

This mathematical model uses a set of intermediate variables $y_k$ which translates into relaxing the upper-bound demand constraints in \citep{jau16_ICTON}. In fact, the constraints \eqref{eq:one_config_per_k_ejor} are a relaxation of the definition \eqref{eq:yk_definition}. As a consequence of this relaxation, the two-variable-sets model requires to post-process the \confs{} in the final solution, as to remove the extra provisioning that violates the demand constraints and to re-establish the equality \eqref{eq:yk_definition}.

Additionally, we have experimented with a model where the intermediate variables are aggregated per node-pair ($\ysd = \sum_{c \in C} \asdc z_c$) such that their number can be lesser. This prospect was unfruitful because the variables $\ysd$ then had to be integer (whereas the variables $y_k$ are binary), thus having a loose domain, which impacts the tightness of the model negatively.

In the next sections, we describe the \NCG{} technique and the solution design and we present the results of the model defined above with intermediate binary variables $y_k$. Prior to that, we present In the sequel of this section a discussion about a possible extension of the mathematical model.

\subsection{Guard-band Extension of the RSA Problem} \label{chap:ApdxA}

When provisioning traffic demand in an \EON{}, a guard band of one slot is reserved on top of the spectrum slice of each request.

If we consider the case where traffic requests are not aggregated, this means that if there are multiple requests for the same node pair, these requests can be provisioned independently, thus, eventually taking different routing paths, and using spectrum slices that are not neighbors. If such requests happen to be routed on the same path using neighbors spectrum slices, they need not to be separated by a guard-band.

In order to account for this possibility, we extend our model as follows:

Given a node pair $w = \{u, v\}$, let $K_w = \{k_1, k_2,..., k_n\}$ be the set of requests between u and v. We will refer to these requests as \textit{atomic}. For $m \in \{1, 2,..., n\}$, we denote ${K'}_w^m$ the set of combinations of m requests from $K_w$. We have 

\begin{equation}
\vert {K'}_w^m \vert = \binom{n}{m}.
\end{equation}

This quantity is the $m^{th}$ binomial coefficient for power $n$. Let ${K'}_w = \cup_{m=1}^{m=n} \> {K'}_w^m$. We have 

\begin{equation}
\vert {K'}_w \vert = \sum_{m=1}^{m=n} \binom{n}{m} = 2^n - 1.
\end{equation}

Each combination of atomic requests produces a new derived request by summing up their demands. ${K'}_w$ is referred to as the set of \textit{derived} requests. Note that $K_w \subset {K'}_w$ given that $K_w = {K'}_w^1$. The derived requests obtained for $m \in \{2,..., n\}$ are referred to as \textit{composite} requests. Let $K' = \cup_{w \in W} {K'}_w$. If a request $k'$ is derived from an atomic request $k$, we note $k' \leftarrow k$.

\begin{figure}[ht]
	\centering
	\includegraphics[width=0.65\textwidth]{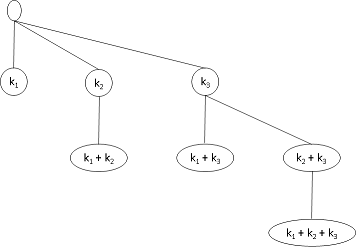}
	\caption{A binomial tree of order n = 3 for a node pair with 3 atomic requests}
	\label{fig:binomtree}
\end{figure}

For each node pair, we enumerate the derived requests using a binomial tree as in \Cref{fig:binomtree}. Given this data structure's properties, this operation is done recursively as we go through the atomic requests of a given node pair. For each derived request, we keep track of the atomic requests used in its making.

In order to account for all possible derived requests, the ILP model is slightly modified as follows, such that only the first set of constraints is impacted:

\begin{equation}
z \quad = \quad \max \quad  \sum_{k \in K} D_k y_k  \label{eq:objective_ejor_ext} 
\end{equation}

subject to:

\begin{alignat}{2}
  &  y_k \leq \sum_{c \in C} \sum_{k' \leftarrow k} a_{k'}^c z_c \qquad    && k \in K, \label{eq:one_config_per_k_ejor_ext}
\\&  \sum_{c \in C} \blsc z_c \leq 1 \qquad     && \ell \in L, s \in S, \label{eq:G_config_disjointness_ejor_ext}
\\&  z_c \in \{0, 1\} \qquad                    && c \in C, \label{eq:domain_zc_ejor_ext}
\\&  0 \leq y_k \leq 1 \qquad                   && k \in K. \label{eq:domain_yk_ext}
\end{alignat}

This modification means that the \PP{} needs to include all the \textit{derived} requests. As calculated earlier, the number of \textit{derived} requests is much larger than the number of \textit{atomic} requests. Therefore, while this extension offers an elegant way to account for non-aggregated connections and saving on the optical spectrum guard-band, by enumerating \textit{derived} requests, it remains unscalable for a practical dataset such as the one introduced in \citep{enoch18_inoc}, and therefore, it is presented for its theoretical value and is not implemented in this study.

\section{Nested Column Generation}
\label{subsec20}

\CG{} consists of decomposing the original problem into a \RMP{} (RMP), i.e., with a restricted number of variables, and one or several \PPs{} (PP). The RMP and the PP(s) are solved alternately. Solving the RMP consists in selecting the best columns, while solving one PP allows the generation of an improving potential column, i.e., a column such that, if added to the current RMP, improves the optimal value of its LP relaxation. 
The process continues until the optimality condition is satisfied, i.e., the so-called reduced cost that defines the objective function of the \PPs{} is non-positive for all of them \citep{chv83}. 
The optimal value of the linear relaxation of the so-obtained RMP is denoted by $\zLPstar$ and represents an upper-bound on the optimal solution of the RSA problem.
After the \CG{} phase, the next step consists of solving the ILP model associated with the last RMP, which produces an integral solution with a value, denoted as $\zILPtilda$, which represents a lower-bound on the optimal solution of the RSA problem.
The integral solution thus-obtained is said to be $\varepsilon$-optimal where $\varepsilon$ denotes the relative difference between the two bounds: $\quad \varepsilon = (\zLPstar - \zILPtilda) / \zLPstar$.

Considering the \conf{} decomposition in \citep{jau16_ICTON}, although \CG{} is a powerful technique, we observed that the link-based formulation of the \PP{} was not efficient enough. Given that the \PPs{} need to be solved at each iteration of \CG{}, their performance can be an important hindrance to the over-all efficiency. In order to address this difficulty, we devised an algorithm based on the idea of solving the \PP{} itself using \CG{}, which approach we referred to as \NCG{}, and which we realized that it has been employed in other optimization problems in the vast area of \textit{Operations Research}.

\subsection{Existing Nested Decompositions}

The idea of applying recursive decomposition was suggested by \cite{dan63}. Some of the first generic implementations for \ILPing{} go back to the early 70's such as \cite{glassey1973nested} and \cite{ho1974nested}. Subsequently, many implementations for \ILPing{} have been produced. 


\cite{vanderbeck2001nested} implemented a nested decomposition approach to a cutting-stock problem. First the author devises a subproblem that generates \textit{cutting patterns} and solves it using \CG{}, in turn, with the help of another subproblem that generates \textit{sections}. The author notes that the cutting-pattern generation subproblem is only solved approximately given that \CG{} only produces lower and upper bounds on the minimum reduced cost of a cutting-pattern, and uses the lower bounds on the reduced costs to produce a Lagrangian bound on the cutting problem. The author recognizes that this is a "heuristic based on the tools of exact optimization", given that the optimality of the Lagrangian bound is not guaranteed.


\cite{hen12} proposed a \NCG{} algorithm for the \textit{Crude Oil Tanker Routing and Scheduling} problem such that the first subproblem generates a cargo-route for each ship. This subproblem is solved using a \textit{Branch-and-Price} algorithm with the help of a second-level subproblem which generates a route for each ship.

Closer to the applications in optical networks, \cite{vignac2016reformulation} presented multiple models for the \textit{Grooming, Routing and Wavelength Assignment} problem, among which, a \CG{} algorithm where a subproblem for each wavelength is defined to find the traffic carried by this wavelength, called a \textit{Wavelength Routing Configuration}. This subproblem is itself decomposed into arc-disjoint grooming-patterns, thus leading to a nested decomposition. Similarly to \cite{vanderbeck2001nested}, the bounds of the \PPs{} are used to compute Lagrangian dual bounds on the master LP, thus resulting into a heuristic.

\cite{caprara2016solving} referred to \NCG{} as \textit{Recursive Dantzig-Wolfe Reformulation} and used this approach to design a \BnP{} algorithm for solving the \textit{Temporal Knapsack Problem}.

\cite{TILK2019549} proposed an algorithm for solving the \textit{Vehicle Routing Problem with Multiple Resource Inter-dependencies} where both the \MP{} and the \PP{} are solved using Branch-and-Price. The authors also give a detailed review on the use of nested decompositions in other publications (e.g., \cite{cord01,DOHN2013157}).

\subsection{Processing Flow}
\label{sec_flow}

\begin{figure}[ht]
	\centering
	\includegraphics[width=1\linewidth]{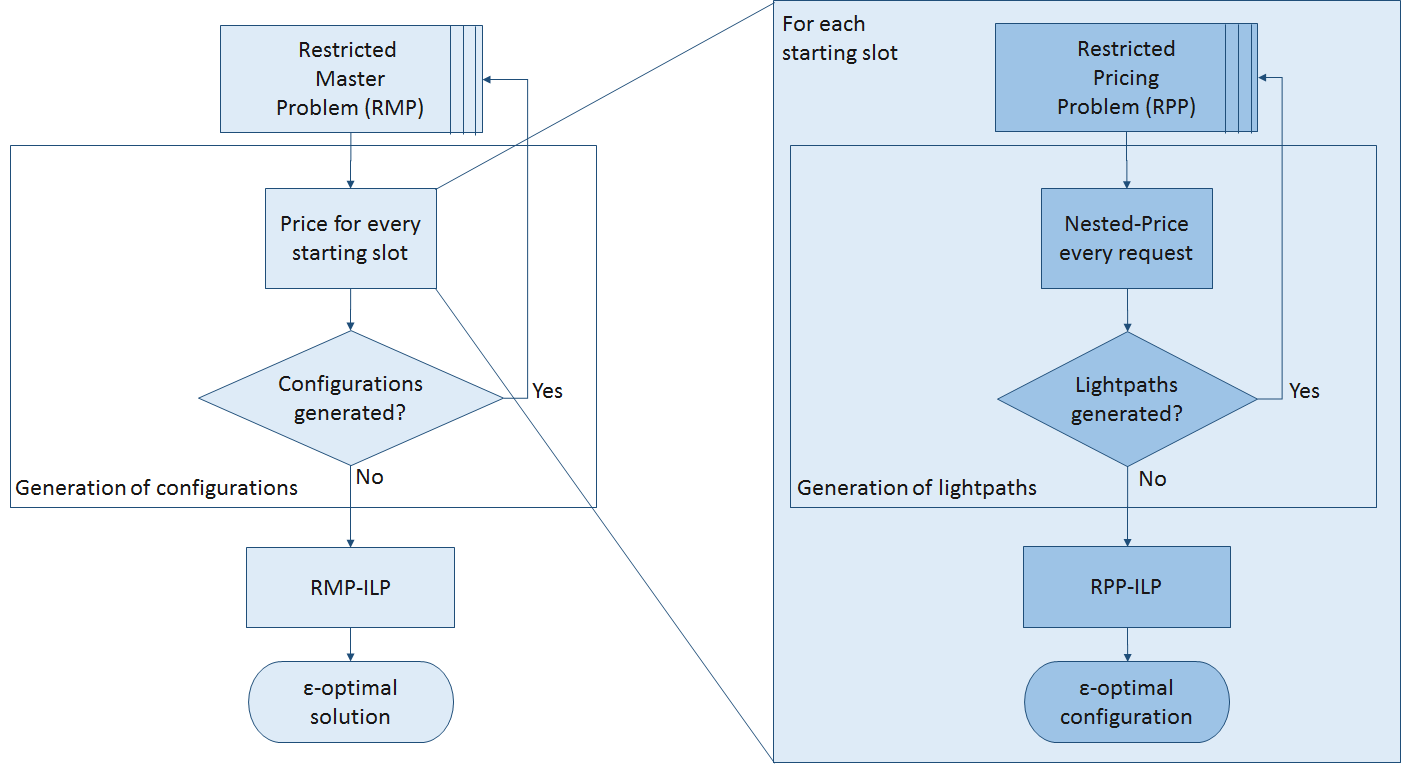}
	\caption{Nested Column Generation Flow for RSA}
	\label{fig:chart_rsa}
\end{figure}

\Cref{fig:chart_rsa} shows the processing flow of our \NCG{} algorithm as applied to the RSA problem. The \CG{} flow used to solve the Pricing problem, is identical to that used to solve the Master problem, thus the naming of \NCG{}. 

The stop condition for the generation of \confs{} for the \MP{} is met when, after considering all the starting slots of the optical spectrum, the \PP{} cannot produce any improving \textit{Configurations}, i.e., the optimal value of the ILP associated with the \PP{} is null ( $ \rcILPstar = 0 $).

\subsection{Discussion about the optimality of the Linear Relaxation of the RMP}
\label{sec_proof}

The main outcome of \CG{} is to produce a strong linear relaxation bound $\zLPstar{}$ on the optimal solution of the \MP{}. This outcome is based on the premise that the \PP{} is solved to optimality \citep{dan63}, in particular, when the stopping condition to exit \CG{} is verified ($ \rcILPstar = 0 $). However, in \NCG{}, the \PP{} is solved using \CG{} meaning that the integer solutions that we obtain for the \PPs{} are only lower-bounds (assuming \BnP{} is not employed as is the case in this study).

Nevertheless, thanks to \NCG{}, we have also an upper bound on the optimal solution of the \PP{} ($\rcLPstar$). So when the \PP{} fails to generate a new \conf{} ($\rcILPtilda = 0$), there are two possibilities:

\begin{itemize}
	\item 
	$\rcLPstar = 0$: we can deduce that $\rcILPstar = 0 $, and therefore the exit condition is satisfied. In this case, the LP relaxation of the \MP{} is optimal.
	
	\item 
	$\rcLPstar > 0$: this means there is no guarantee that $\rcILPstar = 0 $, and therefore, the LP relaxation of the \MP{} might not be optimal.
	
\end{itemize}

To sum up, theoretically \NCG{} does not guarantee the accuracy of the LP relaxation bound of the \MP{}. In fact, some studies resorted to a \BnP{} approach (with additional overhead) for solving the \PP{} in order to guarantee this accuracy \citep{hen12, caprara2016solving,TILK2019549}. However, it is possible empirically to conclude whether the LP relaxation is optimal or not by verifying if the \PP{} was solved to optimality in the last iteration.

\section{Solution Design}
\label{sec2}

%

\subsection{Pricing Problem}
\label{subsec21}

As shown in \Cref{fig:chart_rsa}, the goal of the \PP{} is to compute an optimal \conf{} for each starting slot $s$ (for clear distinction from the starting slot, we will index the provisioning slots by $s'$). In this section we also drop the \conf{} index $c$ in order to alleviate the notation.

Given that the starting slot is fixed for each \PP{}, the optical slice allocated to any request $k$ can be deduced from the starting slot and the size of the request demand $D_k$. Therefore each \PP{} needs only to produce the routing paths for the lightpaths of the optimal \conf{}. Let $\Pi_k$ be the set of all possible routing paths of request $k$. A routing path $\pi \in \Pi_k$ of request $k$ is defined with the help of a binary parameter $\deltapl$ that indicates if link ${\ell}$ is part of the routing path $\pi$.

We define a set of decision variables for the \PP{} as follows:

\begin{itemize}
	\item $\betap$ is a binary variable that indicates if routing path $\pi$ is included in the the optimal \conf{}.
\end{itemize}

Let $\muk{}$ and $\mulsprime{}$ be the dual values of constraints \eqref{eq:one_config_per_k_ejor} and \eqref{eq:G_config_disjointness_ejor} respectively.
The reduced cost for the model \eqref{eq:objective_ejor}-\eqref{eq:domain_yk} is:

\begin{equation}
    rc \quad = \quad \max \qquad \sum_{k \in K} \muk \ak - \sum_{\ell \in L} \sum_{s' \in S} \mulsprime \blsprime .
\end{equation}

The correspondence between the pricing solution and the master's parameters is as follows:

\begin{align}
  & \ak = \sum_{\pi \in \Pi_k} \betap && k \in K
\\& \blsprime = 0 && \ell \in L, s' \notin \{s, \dots, s + D_k^s - 1\}
\\& \blsprime= \sum_{k \in K} \sum_{\pi \in \Pi_k} \deltapl \betap && \ell \in L, s' \in \{s, \dots, s + D_k^s - 1\}.
\end{align}

Therefore, for a given starting slot $s$, the pricing problem reads as follows:

\begin{equation}
	rc \quad = \quad \max 
	  \sum_{k \in K} \sum_{\pi \in \Pi_k} \left( \muk
	 - \sum_{\ell \in L} (\sum_{s'=s}^{s + D_k^s - 1} \mulsprime) \deltapl \right) \betap
\end{equation}

subject to:

\begin{align}
  & \sum_{\pi \in \Pi_k} \betap \leq 1
  && k \in K \label{eq:one_path} 
\\& \sum_{k \in K} \sum_{\pi \in \Pi_k} \deltapl \betap \leq 1
  && \ell \in L \label{eq:path-disjoint} 
\\& \betap \in \{0, 1\}
  && k \in K, \pi \in \Pi_k . \label{eq:domain_beta} 
\end{align}

Constraints \eqref{eq:one_path} mean that each request is provisioned at most once, and constraints \eqref{eq:path-disjoint} mean that each link is traversed by at most one lightpath. This ensures that the lightpaths forming the \conf{} are link-disjoint.

\subsection{Lightpath Generator}
\label{subsec22}

Let $\nuk$ and $\nul$ be the values of the dual variables associated with constraints \eqref{eq:one_path} and \eqref{eq:path-disjoint} respectively. Given a starting slot and given a request $k \in K$, the reduced cost is:

\begin{equation}
\max \quad
     \muk
   - \sum_{\ell \in L} (\sum_{s'=s}^{s + D_k^s - 1} \mulsprime) \delta^\ell
   - \nuk
   - \sum_{\ell \in L} \nul \delta^\ell
\end{equation}

which we streamline as:

\begin{equation}
\max \qquad
     (\muk - \nuk)
   - \sum_{\ell \in L} \left( (\sum_{s'=s}^{s + D_k^s - 1} \mulsprime) + \nul \right) \delta^\ell
\end{equation}

We get a shortest-path problem with non-negative weights for each $k \in K$:
\begin{equation}
\min \qquad \sum_{\ell \in L} \left( (\sum_{s'=s}^{s + D_k^s - 1} \mulsprime) + \nul \right) \delta^\ell.
\end{equation}

\section{Empirical Study}
\label{sec_empiric}

The datasets used in this empirical study use the Spain network topology with 21 nodes and 35 links, and the USA network topology with 24 nodes and 43 links.

The solution was implemented in Java. For the \BnB{} part we used IBM \cplex{} solver 12.6 in a single thread mode. Shortest path calculation was done with an open-source Java library JGraphT implementing Dijkstra's algorithm.
The results were produced on a 3.6-4.0 GHz 4-core machine with 32 GB of memory.

Throughout  the LP phase, only the columns in the basis are retained, and the columns that drop from the basis are dropped from the current RMP, as suggested by \cite{dan63}, in order to maintain the size of the LP manageable, and force the algorithm to produce better columns in terms of low-fractionality.

It is known that linear programs' solvers (\cplex{} in this paper) induce some very small numerical errors. For being so small, usually these errors are not a big hindrance in Column Generation algorithms. But in this study, these errors can be quite a problem. In fact, at the second level of our \NCG{}, the lightpath generator is not solved as a linear program but is solved as an undirected graph shortest-path problem using Dijkstra which does not allow negative link weights. But, while all dual values in this paper are supposed to be non-negative, \cplex{} produces for some of them small negative values. In order to overcome this issue, we are rounding up to zero those negative dual values used in the link weights.

We observe for all the data instances that: When the \PPs{} fail to generate new \confs{} (i.e., $ \rcILPtilda = 0 $), we have also $ \rcLPstar = 0 $, which implies that $\rcILPstar = 0 $ meaning that the \PPs{} are solved optimally. Therefore, the LP relaxation of the \MP{} $\zLPstar $ is optimal and is indeed an upper-bound on the optimal solution $\zILPstar $ (see discussion at \Cref{sec_proof}).

\subsection{ICTON'16 dataset}
\label{dataset_icton}

This dataset was first introduced in \citep{jau16_ICTON} at the conference \textit{ICTON'2016}, where it was solved using a \CG{} algorithm based on a \conf{} decomposition which we refer to as \CCG{}. This dataset was generated using the Spain network topology. \Cref{fig:results_icton} shows the results obtained using the current algorithm as well as a comparison with the results published in \citep{jau16_ICTON} and the results published in \citep{enoch18_inoc}.

\begin{table}[ht]
	\centering
	\caption{ICTON'16 dataset - Solution and Comparison}
	\label{fig:results_icton}
  \resizebox{\textwidth}{!}{%
	\begin{tabular}{rrrrrrrrrrrr}
		\hline
		\multicolumn{3}{|c|}{Traffic demand}                                                                                      & \multicolumn{3}{c|}{Solution quality}                           & \multicolumn{3}{c|}{$\epsilon$ (\%) comparison}                & \multicolumn{3}{c|}{CPU comparison (sec)}                    \\ \hline
		\multicolumn{1}{|r}{\begin{tabular}[c]{@{}r@{}}Offered\\ load (Tbps)\end{tabular}} & \textbar{}D\textbar{}       & \multicolumn{1}{r|}{\textbar{}S\textbar{}} & $\zLPstar$ (Tbps) & $\zILPtilda$ (Tbps) & \multicolumn{1}{r|}{GoS (\%)} & \textbf{C\_NCG} & $\LCGS$        & \multicolumn{1}{r|}{C\_CG} & \textbf{C\_NCG} & $\LCGS$        & \multicolumn{1}{r|}{C\_CG}  \\ \hline
		\multicolumn{12}{|c|}{Spain network: demand granularities in \{1, 2,, 8\} slots, i.e., \{25, 50, ..., 200\} Gbps}                                                                                                                                                                                                        \\ \hline
		\multicolumn{1}{|r}{3.7}                                                           & 35        & \multicolumn{1}{r|}{50}  & 3.7         & 3.6               & \multicolumn{1}{r|}{98.0}     & 2.8             & 0.0          & \multicolumn{1}{r|}{14.0}  & 0.5             & 0.3          & \multicolumn{1}{r|}{50.0}   \\
		\multicolumn{1}{|r}{4.8}                                                           & 45        & \multicolumn{1}{r|}{60}  & 4.8         & 4.7               & \multicolumn{1}{r|}{98.9}     & 1.1             & 0.0          & \multicolumn{1}{r|}{13.0}  & 0.6             & 0.4          & \multicolumn{1}{r|}{86.0}   \\
		\multicolumn{1}{|r}{6.8}                                                           & 60        & \multicolumn{1}{r|}{75}  & 6.8         & 6.72              & \multicolumn{1}{r|}{99.3}     & 0.7             & 0.0          & \multicolumn{1}{r|}{15.0}  & 0.8             & 0.7          & \multicolumn{1}{r|}{147.0}  \\
		\multicolumn{1}{|r}{7.5}                                                           & 64        & \multicolumn{1}{r|}{85}  & 7.5         & 7.15              & \multicolumn{1}{r|}{96.0}     & 4.9             & 0.0          & \multicolumn{1}{r|}{19.0}  & 0.8             & 1.3          & \multicolumn{1}{r|}{176.0}  \\
		\multicolumn{1}{|r}{7.4}                                                           & 70        & \multicolumn{1}{r|}{100} & 7.4         & 7.32              & \multicolumn{1}{r|}{99.3}     & 0.7             & 0.0          & \multicolumn{1}{r|}{16.0}  & 1.2             & 1.7          & \multicolumn{1}{r|}{263.0}  \\
		\multicolumn{1}{|r}{9.7}                                                           & 80        & \multicolumn{1}{r|}{120} & 9.7         & 9.52              & \multicolumn{1}{r|}{98.4}     & 1.6             & 0.0          & \multicolumn{1}{r|}{16.0}  & 1.4             & 2.5          & \multicolumn{1}{r|}{323.0}  \\
		\multicolumn{1}{|r}{12.0}                                                          & 112       & \multicolumn{1}{r|}{150} & 12.0        & 11.95             & \multicolumn{1}{r|}{100.0}    & 0.0             & 0.0          & \multicolumn{1}{r|}{14.0}  & 1.7             & 5.3          & \multicolumn{1}{r|}{417.0}  \\
		\multicolumn{1}{|r}{20.5}                                                          & 180       & \multicolumn{1}{r|}{330} & 20.5        & 20.2              & \multicolumn{1}{r|}{98.4}     & 1.6             & 0.0          & \multicolumn{1}{r|}{18.0}  & 4.0             & 25.6         & \multicolumn{1}{r|}{1606.0} \\ \hline
		\multicolumn{12}{|c|}{Spain network: demand granularities in \{2, 4,, 16\} slots, i.e., \{50, 100, ..., 400\} Gbps}                                                                                                                                                                                                      \\ \hline
		\multicolumn{1}{|r}{7.5}                                                           & 35        & \multicolumn{1}{r|}{80}  & 7.5         & 7.4               & \multicolumn{1}{r|}{99.3}     & 0.7             & 0.0          & \multicolumn{1}{r|}{10.0}  & 0.8             & 0.9          & \multicolumn{1}{r|}{134.0}  \\
		\multicolumn{1}{|r}{9.8}                                                           & 45        & \multicolumn{1}{r|}{110} & 9.8         & 9.7               & \multicolumn{1}{r|}{99.5}     & 0.5             & 0.0          & \multicolumn{1}{r|}{10.0}  & 0.9             & 2.0          & \multicolumn{1}{r|}{177.0}  \\
		\multicolumn{1}{|r}{10.7}                                                          & 60        & \multicolumn{1}{r|}{156} & 10.7        & 10.65             & \multicolumn{1}{r|}{99.5}     & 0.5             & 0.0          & \multicolumn{1}{r|}{12.0}  & 0.9             & 3.1          & \multicolumn{1}{r|}{261.0}  \\
		\multicolumn{1}{|r}{15.5}                                                          & 64        & \multicolumn{1}{r|}{170} & 15.5        & 15.5              & \multicolumn{1}{r|}{100.0}    & 0.0             & 0.0          & \multicolumn{1}{r|}{16.0}  & 1.4             & 4.7          & \multicolumn{1}{r|}{630.0}  \\
		\multicolumn{1}{|r}{15.1}                                                          & 70        & \multicolumn{1}{r|}{236} & 15.1        & 15.1              & \multicolumn{1}{r|}{100.0}    & 0.0             & 0.0          & \multicolumn{1}{r|}{13.0}  & 1.4             & 7.8          & \multicolumn{1}{r|}{1342.0} \\
		\multicolumn{1}{|r}{16.9}                                                          & 80        & \multicolumn{1}{r|}{256} & 16.9        & 16.85             & \multicolumn{1}{r|}{100.0}    & 0.0             & 0.0          & \multicolumn{1}{r|}{14.0}  & 1.4             & 10.3         & \multicolumn{1}{r|}{1419.0} \\ \hline
		\textbf{Average}                                                                   & \textbf{} & \textbf{}                & \textbf{}   & \textbf{}         & \textbf{}                     & \textbf{1.1}    & \textbf{0.0} & \textbf{14.3}              & \textbf{1.3}    & \textbf{4.8} & \textbf{502.2}             
	\end{tabular}%
}
\end{table}

We did not re-implement the algorithm \CCG{} of \cite{jau16_ICTON} and the comparison is made with the results that were published by the authors of that algorithm. Considering that they produced their results on a 1.9-2.5GHz machine (in contrast with our 3.6-4.0GHz machine), it would be reasonable to divide their computation times by two for the purpose of the comparison (below) that remains approximate.

While the \CCG{} algorithm by \cite{jau16_ICTON} takes around 4 minutes on average, the new \CNCG{} algorithm takes only 1.3 seconds on average, beating even the \LCGS{} algorithm by \cite{enoch18_inoc} at around 5 seconds on average.

Regarding the quality of the solution, while the \LCGS{} algorithm solves this dataset to optimality, the \CNCG{} algorithm produces a solution accuracy around 1\% on average, which is still a lot better than the accuracy of the \CCG{} algorithm of 14.3\%. 

In the next subsection, we will examine a dataset that is supposed to be much harder, and shall be more challenging for the new \CNCG{} algorithm.

\subsection{INOC'18 dataset}
\label{dataset_iNOC}

This dataset was first introduced in \citep{enoch18_inoc}, where it was solved using a \CG{} algorithm \LCGS{} based on a \lp{} decomposition. This dataset was generated using both the Spain network and the USA network topologies according to the following specifications:

\begin{itemize}
	\item Demands are drawn from granularities \{4, 8, 16\} in units of spectrum slots, with respective proportions \{70\%, 20\%, 10\%\}, thus generating requests that are relatively large compared to those encountered in today's networks, hence producing instances with relatively a high level of difficulty.

	\item Offered load is spread over all node pairs. Therefore, after aggregation, the number of requests given as input to our algorithm corresponds to the number of node pairs in the network.

\end{itemize}

In order to ease down the complexity of the algorithm and enhance its execution, we opted for the following strategies:

\begin{itemize}

    \item Requests are aggregated per node-pair. This has the advantage of reducing the number of requests in the optimization problem and easing-down the solving process. Note that theoretically, if the problem is solved to optimality, this strategy would potentially produce a solution with a worse quality than if the requests were not aggregated.

    \item Given the large size of the instances of this dataset, the numerical accuracy of the LP solving becomes crucial in order for the \CG{} to converge. To that end, \cplex{} parameter \textit{NumericalEmphasis} \footnote{Numerical precision emphasis} is turned on.

	\item We have noticed that the \ILP{} phase can take a long time to cover the whole of the \BnB{} tree, without improving by much the integral solution. Therefore we have chosen to pre-terminate the \BnB{} by setting the \cplex{} parameter \textit{EpGap} \footnote{Relative MIP gap tolerance: default: 1e-04.} to $10^{-1}$.
	
	\item We allow multi-threading during the \BnB{} phase with up to 8 threads.
	
	\item Before the \ILP{} phase, we set \cplex{} parameter \textit{Advance} \footnote{If set to 1 or 2, this parameter specifies that CPLEX should use advanced starting information when it initiates optimization.} to 0 in order to enable the standard pre-solving operations instead of using the basis produced by the last iteration of \CG{} as a starting point for the \BnB{}.

\end{itemize}

\Cref{fig:results_inoc} shows the complete results obtained using the current algorithm in terms of the $\epsilon$-optimal solution as well as the computation time. We observe that instances Spain90, USA80 and USA90 register a spike in the computation time as compared to the other instances. As we have discussed in \citep{enoch18_inoc}, this is due to the fact that $\zLPstar{}$ is lesser than the offered load. Consequently, the \CG{} needs a considerably larger number of columns to reach the $\zLPstar{}$, and spends a relatively long time producing those columns. This dataset has been purposefully designed to showcase this type of difficult instances.

\begin{table}[ht]
	\centering
	\caption{INOC'18 dataset - Algorithm Results and Performance}
	\label{fig:results_inoc}
  \resizebox{\textwidth}{!}{%
	\begin{tabular}{|lrrr|rrrr|rrr|}
		\hline
		\multicolumn{4}{|c|}{Dataset}       & \multicolumn{4}{c|}{Solution quality}               & \multicolumn{3}{c|}{CPU time (sec)} \\ \hline
		Instance  & \textbar{}S\textbar{} & \textbar{}D\textbar{} & Load (Tbps) & $\zLPstar$ (Tbps) & $\zILPtilda$ (Tbps) & $\epsilon$ (\%) & GoS (\%) & LP          & ILP      & Total      \\ 
		\hline
		\multicolumn{11}{|c|}{Spain network: Demands in \{4,8, 16\} slots, i.e., \{100, 200, 400\} Gbps}                                \\ \hline
		Spain\_50 & 400 & 413 & 50.2        & 50.2        & 42.6              & 17.8   & 84.9     & 30.5        & 0.3      & 31.0       \\
		Spain\_60 & 400 & 495 & 60.0        & 60.0        & 47.2              & 27.1   & 78.7     & 113.6       & 0.2      & 114.1      \\
		Spain\_70 & 400 & 578 & 70.2        & 70.2        & 52.3              & 34.2   & 74.5     & 177.0       & 1.0      & 178.2      \\
		Spain\_80 & 400 & 660 & 80.0        & 80.0        & 55.5              & 44.1   & 69.4     & 254.1       & 2.2      & 256.6      \\
		Spain\_90 & 400 & 743 & \textbf{90.2}        & \textbf{86.9}        & 56.0              & 55.1   & 62.1     & 1,756.9     & 940.1    & 2,697.7    \\ \hline
		\multicolumn{11}{|c|}{USA network: Demands in \{4,8, 16\} slots, i.e., \{100, 200, 400\} Gbps}                                  \\ \hline
		USA\_50   & 400 & 413 & 50.2        & 50.2        & 41.4              & 21.3   & 82.5     & 27.8        & 0.5      & 28.4       \\
		USA\_60   & 400 & 495 & 60.0        & 60.0        & 45.6              & 31.6   & 76.0     & 350.4       & 3.1      & 353.8      \\
		USA\_70   & 400 & 578 & 70.2        & 70.2        & 49.9              & 40.7   & 71.1     & 495.4       & 3.1      & 498.9      \\
		USA\_80   & 400 & 660 & \textbf{80.0}        & \textbf{78.3}        & 53.2              & 47.2   & 66.5     & 1,441.2     & 204.2    & 1,646.3    \\
		USA\_90   & 400 & 743 & \textbf{90.2}        & \textbf{85.7}        & 56.0              & 53.0   & 62.1     & 1,977.5     & 417.0    & 2,395.3    \\ \hline
	\end{tabular}%
}
\end{table}

\Cref{fig:comparison_inoc} shows a comparison with the results published in \citep{enoch18_inoc}. We can make the following observations:

\begin{table}[ht]
	\centering
	\caption{INOC'18 dataset - Solution and Performance Comparison}
	\label{fig:comparison_inoc}
  \resizebox{\textwidth}{!}{%
	\begin{tabular}{lrrrrrrrrr}
		\hline
		\multicolumn{2}{|c|}{Dataset}                                     & \multicolumn{6}{c|}{Quality comparison}                                                                                            & \multicolumn{2}{c|}{\multirow{2}{*}{\begin{tabular}[c]{@{}c@{}}CPU comparison\\ (sec)\end{tabular}}} \\ \cline{1-8}
		\multicolumn{1}{|c}{}          & \multicolumn{1}{c|}{}            & \multicolumn{2}{c}{$\zLPstar$ (Tbps)}           & \multicolumn{2}{c}{$\zILPtilda$ (Tbps)}           & \multicolumn{2}{c|}{$\epsilon$ (\%)}                 & \multicolumn{2}{c|}{}                                                                                \\ \cline{3-10} 
		\multicolumn{1}{|l}{Instance}  & \multicolumn{1}{r|}{Load (Tbps)} & \textbf{$\CNCG$} & \multicolumn{1}{r|}{$\LCGS$} & \textbf{$\CNCG$} & \multicolumn{1}{r|}{$\LCGS$} & \textbf{$\CNCG$}  & \multicolumn{1}{r|}{$\LCGS$} & \textbf{$\CNCG$}                               & \multicolumn{1}{r|}{$\LCGS$}                              \\ \hline
		\multicolumn{1}{|l}{Spain\_50} & \multicolumn{1}{r|}{50.2}        & 50.2         & \multicolumn{1}{r|}{50.2}  & 42.6         & \multicolumn{1}{r|}{48.0}  & 17.8          & \multicolumn{1}{r|}{4.6}   & 31.0                                       & \multicolumn{1}{r|}{22.4}                               \\
		\multicolumn{1}{|l}{Spain\_60} & \multicolumn{1}{r|}{60.0}        & 60.0         & \multicolumn{1}{r|}{60.0}  & 47.2         & \multicolumn{1}{r|}{57.8}  & 27.1          & \multicolumn{1}{r|}{3.8}   & 114.1                                      & \multicolumn{1}{r|}{28.1}                               \\
		\multicolumn{1}{|l}{Spain\_70} & \multicolumn{1}{r|}{70.2}        & 70.2         & \multicolumn{1}{r|}{70.2}  & 52.3         & \multicolumn{1}{r|}{66.6}  & 34.2          & \multicolumn{1}{r|}{5.4}   & 178.2                                      & \multicolumn{1}{r|}{42.2}                               \\
		\multicolumn{1}{|l}{Spain\_80} & \multicolumn{1}{r|}{80.0}        & 80.0         & \multicolumn{1}{r|}{80.0}  & 55.5         & \multicolumn{1}{r|}{73.6}  & 44.1          & \multicolumn{1}{r|}{8.7}   & 256.6                                      & \multicolumn{1}{r|}{71.9}                               \\
		\multicolumn{1}{|l}{Spain\_90} & \multicolumn{1}{r|}{\textbf{90.2}}        & \textbf{86.9}         & \multicolumn{1}{r|}{86.9}  & 56.0         & \multicolumn{1}{r|}{75.0}  & 55.1          & \multicolumn{1}{r|}{15.8}  & 2,697.7                                    & \multicolumn{1}{r|}{1,131.1}                            \\
		\hline
		\multicolumn{1}{|l}{USA\_50}   & \multicolumn{1}{r|}{50.2}        & 50.2         & \multicolumn{1}{r|}{50.2}  & 41.4         & \multicolumn{1}{r|}{46.9}  & 21.3          & \multicolumn{1}{r|}{7.0}   & 28.4                                       & \multicolumn{1}{r|}{29.4}                               \\
		\multicolumn{1}{|l}{USA\_60}   & \multicolumn{1}{r|}{60.0}        & 60.0         & \multicolumn{1}{r|}{60.0}  & 45.6         & \multicolumn{1}{r|}{55.0}  & 31.6          & \multicolumn{1}{r|}{9.1}   & 353.8                                      & \multicolumn{1}{r|}{55.8}                               \\
		\multicolumn{1}{|l}{USA\_70}   & \multicolumn{1}{r|}{70.2}        & 70.2         & \multicolumn{1}{r|}{70.2}  & 49.9         & \multicolumn{1}{r|}{63.0}  & 40.7          & \multicolumn{1}{r|}{9.9}   & 498.9                                      & \multicolumn{1}{r|}{108.4}                              \\
		\multicolumn{1}{|l}{USA\_80}   & \multicolumn{1}{r|}{\textbf{80.0}}        & \textbf{78.3}         & \multicolumn{1}{r|}{78.3}  & 53.2         & \multicolumn{1}{r|}{65.5}  & 47.2          & \multicolumn{1}{r|}{19.6}  & 1,646.3                                    & \multicolumn{1}{r|}{1,122.0}                            \\
		\multicolumn{1}{|l}{USA\_90}   & \multicolumn{1}{r|}{\textbf{90.2}}        & \textbf{85.7}         & \multicolumn{1}{r|}{85.7}  & 56.0         & \multicolumn{1}{r|}{72.4}  & 53.0          & \multicolumn{1}{r|}{18.4}  & 2,395.3                                    & \multicolumn{1}{r|}{1,532.0}                            \\ 
		\hline
		\textbf{Average}               & \textbf{}                        & \textbf{}    & \textbf{}                  & \textbf{}    & \textbf{}                  & \textbf{37.2} & \textbf{10.2}              & \textbf{820.0}                             & \textbf{414.3}                                         
	\end{tabular}%
}
\end{table}

\begin{itemize}

    \item The upper-bound $\zLPstar{}$ produced by the two algorithms is the same. A theoretical study of this upper-bound could be considered in a future work.

    \item The integral solution found by \CNCG{} algorithm is less good than the one found by the \LCGS{} algorithm and the accuracy gap is a lot larger. This is due to the fact that a \conf{} column is a lot larger than a \lp{} column. Consequently, in the ILP phase, if the \conf{} column has even a small overlapping with another column in the integral solution, it is rejected as a whole. To get an intuition of this phenomenon, we could make an analogy with the \textit{Knapsack} problem. If we suppose that we have a fractional \textit{Knapsack} solution and we want to get the integral solution from it: The smaller the items in the fractional solution, the lesser the contention among them, and the better the integral solution.

    \item The computation time of the \CNCG{} algorithm is less good than that of the \LCGS{} algorithm (double on average). This is also due to the disparity of size of the columns in each decomposition. That said, the computation time of the \CNCG{} algorithm remains reasonable considering the level of difficulty of this dataset.
    
\end{itemize}

While the \CNCG{} algorithm is more scalable than the previous study, it can be inefficient in terms of solution quality for the largest data instances. We conclude that the algorithm \LCGS{}, based on the \lp{} decomposition, presented in \citep{enoch18_inoc} remains the most efficient in terms of both solution quality and scalability.

\section{Conclusion} \label{chap:conclusion}

Considering the provisioning of Elastic Optical Networks, we focused on the \RSA{} (RSA) problem. We aimed to explore ways to improve the latest results published for this problem in \citep{jau16_ICTON}. In doing so, we independently rediscovered the \NCG{} technique, which we used to design an ILP algorithm based on the \conf{} decomposition. The motivation behind this approach is to tame the difficulty of the \PP{} thus improving the overall performance. Additionally, we presented for the first time a theoretical extension of the mathematical model, where we don't assume that the connections are aggregated for each node pair, and we make savings on the optical spectrum guard-band. Our new approach based on \NCG{} produced solutions with an accuracy near optimal for the reference dataset and improved the computation time by a factor of two orders of magnitude on average as compared to the latest existing \conf{} decomposition algorithm in \citep{jau16_ICTON}, which is a considerable leap in efficiency. In addition to the reference dataset used in \citep{jau16_ICTON}, we also produced results for the dataset used in \citep{enoch18_inoc}, consisting of problem instances considerably larger and more difficult to solve. 

Equipped with the \NCG{} technique, we could consider increasing the scope of the RSA problem in the future by adding other dimensions that are useful for real-life implementations such as varying the modulation formats, or minimizing the number of optical re-generators needed in the optical network.

\bibliography{main}

\end{document}